\documentclass[usegraphicx,usenatbib]{mn2e}
\title[Polytropic Supernovae: 1-D Hydrodynamics Calculation]{Explosive Disruption of Polytropes: a One Dimensional Hydrodynamic Calculation}
\author[M. Wyman et al.]{Mark C. Wyman$^{1,2}$\thanks{E-mail: wyman@astro.cornell.edu; chernoff@astro.cornell.edu; ira@astro.cornell.edu}, David F. Chernoff$^{1}$\footnotemark[1], and Ira Wasserman$^{1,2}$\footnotemark[1] \\ $^1$ Center for Radiophysics and Space Research, Cornell University,
Ithaca, NY 14853, USA \\
$^2$ Laboratory for Elementary Particle Physics, Cornell University,
Ithaca, NY 14853, USA}

\def\be{\begin{equation}}
\def\ee{\end{equation}}
\def\baray{\begin{eqnarray}}
\def\earay{\end{eqnarray}}
\def\rstar{R_\star}
\def\mstar{M_\star}
\def\mhat{\hat m}
\def\mdim{\tilde m}
\def\avis{a_{\rm visc}}
\def\qvis{q_{\rm visc}}
\def\ahatvis{\hat a_{\rm visc}}
\def\qhatvis{\hat\qvis}
\def\vhat{\hat v}
\def\Vhat{\hat V}
\def\rhat{\hat r}

\def\phat{\hat P}
\def\uhat{\hat U}
\def\rhohat{\hat\rho}
\def\ehat{\hat E}
 \def\be{\begin{equation}}
 \def\ee{\end{equation}}
 \def\ba{\begin{eqnarray}}
 \def\ea{\end{eqnarray}}
\begin{document}
\date{Submitted 2004 April 23; Resubmitted 2004 July 19}

\maketitle
\label{firstpage}
\begin{abstract}
We study explosions of stellar models using a one-dimensional Lagrangian
hydrodynamics code. We calculate how much mass is liberated as a
function of the energy of explosion for a variety of pre-explosion
polytropic structures and for equations of state with a range of
radiation-to-gas pressure ratios. The results show that simple
assumptions about the amount of mass lost in an explosion can be quite
inaccurate, and that even one-dimensional stellar models exhibit a rich
phenomenology. The mass loss fraction rises from about 50 to 100 per cent as
a function of the explosion energy in an approximately discontinuous
manner. Combining our results with those of other, more realistic models, we suggest that Nova Scorpii (J1655-40) may have experienced significant mass fallback because the explosion energy was less than the critical value.  We infer that the original progenitor was less
than twice the mass of today's remnant.

\end{abstract}
\section{Introduction}

A fundamental question in the study of supernovae is the
fate of a star subject to an explosion of a given strength: 
is the star completely disrupted, 
and, if not, how much of the star is
lost and what is the configuration of the matter that remains bound?
Many researchers have addressed this question for specific cases of
interest using detailed numerical simulations. To our knowledge, a
precise quantitative relationship between the strength of the
explosion and the fate of the outer layers has not been given
before, even for highly idealized stellar models. The
potential utility of such a relationship is evident 
in the analysis of \citet{fry01},
where a simple ``rule of thumb,'' introduced to estimate 
the amount of mass left bound in a supernova, 
allows a determination of which high mass
stars leave behind neutron stars and which ones yield
black holes. The ``rule of thumb'' stipulates
 that a portion (between 30 and 50 per cent) of the explosion energy is effective in
directly unbinding the outermost layers of a
star; this estimate is based on 
detailed simulations by \citet{mac01} for a set of specific stellar progenitors.  

Our goal is to improve the
understanding of the disruption process by carrying out hydrodynamical
calculations of simple, polytropic stellar models with a range of explosion
strengths, polytropic indices, and equations of state. 
We intend this sort of calculation to complement, not replace, 
the realistic, detailed simulations that are the current state-of-the-art in this field. 
Physical complications such as density jumps, neutrino transport, and aspherical motions
are absent from our calculations.
Instead, our goal is to  incorporate the essential physics -- hydrodynamics and gravity -- in 
models that are easy to compute and useful to the study 
of supernovae in the same way that the polytrope itself is useful to 
stellar modeling. We note that this subject has been treated before, 
in a very different perturbative calculation \citep{nad63}.
Already, our results yield improved versions of the 
``rule of thumb,'' which we provide in a simple, easily applied, empirical form. 
Although a host of significant core-collapse 
modeling uncertainties remain (hydrodynamic
motions in the core, distribution of angular momentum within the
collapsing object, neutrino-matter coupling, etc.),
our simplified treatment represents an
improvement in the determination of the fate of central
remnants -- providing a convenient bridge between estimates of mass loss
based upon simple physical assumptions and more sophisticated, realistic simulations.

There is considerable evidence that a supernova
explosion occurred in J1655-40: the atmosphere of its companion is
contaminated with elements thought to be formed only in supernovae
\citep{isr99}, and it is likely that the black hole progenitor was
considerably more massive than the remnant we see today
\citep{oro97,sha99}. There is also some evidence that the J1655-40
system could have remained bound only if it received a substantial
kick during or shortly after the formation of its black hole
\citep{mir02}. While our current models are too simple to provide 
definitive results for any particular system, we
believe that the methods employed here suggest that the progenitor mass was
less than twice the mass of today's remnant.  
Mass fallback may trigger the collapse to a black hole as well as 
pollute the companion's atmosphere. 

In section 2, we describe the physical set up,
while section 3 describes the numerical code. In section 4, we give
 more detailed results and discuss how the numerical data were analyzed, 
 and in section 5 we comment on our results' lack of dependence on 
 the inner boundary condition.

\section{Problem and parameter ranges}

We model the supernova as a spherically symmetric explosion in a stellar model that is
initially in hydrostatic equilibrium. The pre-explosion stellar
structure is polytropic. We deposit the full energy of the explosion
in a small region near the centre of the polytrope. Using a finite-difference
code we calculate the hydrodynamical evolution. A shock propagates towards
the surface and the outer layers of the stellar model may be ejected. If it
is not completely destroyed, part of the model remains
gravitationally bound and we follow the evolution long enough to make
an accurate estimate of the mass of the remnant.

We considered a {\it range of initial stellar structures}. We varied the
polytropic index $\Gamma$ where $P \propto \rho^\Gamma$.  The
Lane-Emden equation prescribes the run of density and pressure in the
initial model; our choices for $n=1/(\Gamma-1)$ span $3/2 \le n \le
4$. As is well-known, the polytrope's ratio of central to mean density
increases as $n$ varies from $0$ to $5$. This range
subsumes typical main-sequence profiles and extended red-giant
structures.

We considered {\it two equation of state treatments}: ideal gas
pressure (``EOS M'': $P = P_{matter}$ with a fixed ratio of specific
heats $\gamma$) and a mixture of gas plus radiation in thermal
equilibrium (``EOS MR'': $P = P_{rad} + P_{matter}$). EOS M is
suitable for stellar models of low mass (dominated by particle pressure at
their centres) and weak explosions (such that the post-shock gas is
not radiation dominated); EOS MR is needed if there is significant
radiation pressure.  We infer the temperature profile from the
appropriate EOS and the Lane-Emden pressure-density profile.  
For EOS MR, we chose to limit ourselves to {\it convectively stable polytropic models.} 
We will discuss the condition for stability in \S 4.2. 
To facilitate the description of our problem's parameter space, 
let $s_c = P_{rad}(r = 0) / P_{matter}(r = 0)$. 
As we will show in \S 4.1, for a given polytropic index $n$, the choice of
$s_c$ uniquely fixes the mass of the resulting stellar model. 
We define our mass scale to be $M_{scale} = M_{\odot} (m_p / 2\mu)^2$.
We then chose the dimensionless masses of the stellar models we exploded to be
 $\mdim = M_{star}/M_{scale} = 10,\; 100,\; 1000$ for a range of ($n$, $s_c$) pairs,
 taking care to remain in the convectively stable region of parameter space. 
 See Tab.\ref{tab:starmasses} and Fig.\ref{fig:stabstars}. 

We considered a {\it variety of explosion energies}. Given our interest
in studying explosions which only partially unbind the stellar models,
 we typically considered blasts with $0.1 \le
E_{\rm blast}/E_{\rm bind} \le 1.5$, \emph{i.e.} energies of the same order of
magnitude as a simple dimensional estimate for unbinding. 

 In brief, the results we obtained were as follows. \begin{itemize}
\item The amount of mass lost as a function of explosion energy
 makes a discrete jump from approximately 50 to 100 per cent in all models,
 suggesting a point of instability.
\item Explosions that do not totally disrupt a stellar model give rise to mass
 loss curves that, in most cases, scale quadratically with explosion energy.
\end{itemize}

These simulations did not
include any sort of compact object at the core
 of our explosions. We explored
the sensitivity of the mass loss to {\it our treatment of the inner boundary condition}.
We found that to a large extent the results are unchanged: \begin{itemize}
\item When two extreme inner boundary conditions -- a `vacuum cleaner'
core that sucks up any incident material, and its opposite,
a hard, reflecting shell -- were compared, the mass loss results were virtually 
identical.
\end{itemize} Readers primarily interested in further details regarding these results are encouraged
to skip to \S 4.

\section{The code and numerical tests}

\subsection{Equations}

We use the inviscid fluid equations which describe mass, momentum and
energy conservation. All calculations are one-dimensional with either
a plane-parallel (for testing) or spherical (for testing and
simulations) geometry. We advance the fluid state using a finite
difference approximation to the fluid equations (Lax-Wendroff,
explicitly differenced, 1-D Lagrangian code [\citealt{ric67}]). Shocks are
handled with the addition of artificial viscosity. We solve Poisson's
equation to determine the gravitational forces at each time step.
Details are provided in Appendices A - C.

\subsection{Tests of hydrodynamics}

We tested the purely hydrodynamic capabilities of the code (no
gravity) via comparison with the Sod shock tube (plane-parallel geometry) and
Sedov blast (spherical geometry) solutions.  For the Sod test with
$\gamma=7/5$ (as well as for a range of other $\gamma$'s), EOS M,
various overpressures ($p_2/p_1 = 10, 100, 1000$), and
$1200$ zones, we found essentially perfect agreement between the numerical and analytic
solutions, except for the shock smearing over $\sim 5-8$ zones.

For the Sedov problem, we re-derived the solution given in {\it Fluid Mechanics} by \citet{ll87},
 thereby finding the correction to that solution's typographical error (in an exponent) 
mentioned in Shu's {\it Gas Dynamics}. The correction 
is recorded in Appendix D. We carried out a number
of blast wave simulations, varying our choices of EOS and
$\gamma$.  For flows dominated by particle pressure we compared
numerical solutions ($\gamma = 5/3$ and $7/5$ for EOS M) with the
analytic similarity solution; for
flows dominated by radiation pressure we compared several different
radiation-dominated numerical solutions ($\gamma=5/3$, EOS MR) to the
$\gamma=4/3$ similarity solution.  The radiation-dominated numerical
solutions were generated by explosions yielding
high-Mach-number shocks.  A range of initial radiation-to-matter pressure ratios
$s$ and explosion energies were considered. One simulation with large
 constant $s\sim 1000$ and
relatively small explosion energy and another simulation with small constant
$s\sim 0.1$ and large energy both yielded a radiation-dominated
post-shock flows.

In all calculations, the explosion was allowed to expand to well over 100 times the size of the initial
``bomb zone." Comparisons of EOS M runs with analytic
solutions were possible throughout the simulation; comparisons of
EOS MR runs with the analytic radiation-dominated $\gamma=4/3$ similarity 
solution were meaningful only for the part of the simulation in which
radiation pressure dominated matter pressure, approximately 4-5
expansion times.  With $800$ zones, the Sedov test gave close
($2-3$ per cent) agreement in the relative density, velocity and pressure of
the numerical solution and the analytic similarity solution for both
particle pressure dominated and radiation pressure dominated flows
except in the central-most region.

Two factors contribute to the discrepancies at the centre. First, the
innermost zone was treated as an adiabatic
expanding/contracting bubble. The entropy of this zone was incorrect
but its mass was so small that its impact on the rest of the solution
was inconsequential. The explosion results were found to be almost
entirely insensitive to alternative methods of treating this innermost zone, 
provided the treatments were energy conserving. Second, explosive energy was
injected in a small, but non-negligible central region (typically the
inner 5 per cent of the mass). Quantitative differences between the analytic
similarity solution and the numerical solution occurred
in the part of the grid used as the ``bomb zone'' and persisted through the
simulation.  These differences were not unexpected, as the point-like nature
of the explosion in the similarity solution cannot be realized in any finite simulation.
We also compared two models for the energy injection at the centre. 
In one, the ``thermal bomb,'' an excess of
thermal energy equal to the desired explosion energy was added by hand
to the core (inner 5 per cent of the mass) of the stellar model, essentially creating
an out-of-equilibrium hot core that then expanded rapidly into the
stellar envelope. In the other, the ``kinetic bomb,'' a linear
velocity profile carrying the same amount of energy was added to the
inner 5 per cent of the mass.  Both methods produced identical results
outside the ``bomb zone.''

\subsection{Tests of hydrostatics}

With the inclusion of self-gravity forces, we verified that
Runge-Kutta integration of the Lane-Emden equations yielded
stationary, stable configurations for our time-dependent hydrodynamic
evolution equations (finite difference scheme). We checked the
long-lived stability for all polytropic indices and radiation-to-gas
pressure ratios adopted in this study. Likewise, we tested that
the virial theorem was satisfied by the initial configurations.

\subsection{Tests of self-gravitating explosions}

In the actual runs of the problem of interest, we further verified
that the treatment of the central zone made no discernible difference,
that variations in the size of the ``bomb zone" (3-10 per cent, for instance), caused
only very slight ($< 5$ per cent) changes to the amount of mass lost in the
explosions. We also verified that energy conservation was satisfied (to $<5$ per cent).

\section{Results}

We adopted polytropes for the initial stellar structure with $P = k
\rho^\Gamma$. The density and pressure profiles were determined by
solving the Lane-Emden equation with the total mass and radius
scaled to unity. We refer to this as the dimensionless solution; it
depends only upon $\Gamma$. The dimensionless density-pressure
distributions are the forms used in our computations. All results are
likewise reported using dimensionless quantities (explosion energy
in terms of binding energy, mass loss in terms of the total mass, etc.). 

\subsection{Scaling of polytropes}

Let us first review the scaling of the initial polytropic solution. For
given $k$ and 
$\Gamma\equiv 1+1/n$ in the pressure-density relation, 
it is possible to generate a one-parameter
family of scaled solutions with
\begin{equation}
M^{2-\Gamma}R^{3\Gamma-4}=(k/G)f(\Gamma)~,
\label{mreqn}
\end{equation}
with $f(\Gamma)$ a dimensionless number depending on polytropic index.
We can construct a polytropic progenitor without radiation 
pressure, but that is an idealization that is approximately
correct only in the limit of a low ratio of radiation to gas pressure.
If the matter has an ideal gas equation of state, $P=\rho kT/\mu$, there
must be nonzero temperature inside the stellar model, and hence nonzero
radiation pressure $P_{\rm rad}=aT^4$. Under some circumstances, $P_{\rm rad}$
will be low both in the progenitor and in the ejecta after the stellar model
explodes. The explosions of such models have a universal mass loss
fraction as a function of $\Gamma$ and explosion energy in units of the stellar
binding energy.

Imposing a fixed value
of the radiation-to-gas pressure $s(r=0) = s_c$ at the centre of the 
stellar model before the explosion reduces the scaling of the dimensionless solution.
These models also have a universal mass loss fraction
as a function of $\Gamma,$ $s_c,$ and explosion energy. The reason the scaling is reduced is 
because specifying $s_c$ determines the stellar mass for a given value
of $\Gamma$ independent of $k$: 
\begin{equation}
M=m(\Gamma)M_{\rm Ch}s_c^{1/2}\left(1+s_c\right)^{3/2}~,
\label{eddington}
\end{equation}
where $M_{\rm Ch}=(\hbar c/G)^{3/2}\mu^{-2}=1.86(m_p/\mu)^2M_\odot = 7.44 M_{scale}$, 
and $m(\Gamma)=
(45/\pi^2)^{1/2}[f_\rho(\Gamma)]^2[f_p(\Gamma)]^{-3/2}$, with
$f_P(\Gamma)=P_cR^4/GM^2$ and $f_\rho(\Gamma)=\rho_cR^3/M$, which
are both dimensionless functions of polytropic index only.
For $n=3/2$, $f_\rho=1.430$ and $f_p=0.7702$, so $m(5/3)=6.460$;
for $n=3$, $f_\rho=12.94$, $f_p=11.05$, so $m(4/3)=9.734$. 
In the low mass limit, this implies $s_c\propto M^2$.
Thus, from Eq. (\ref{mreqn}), the combination
$kR^{4-3\Gamma}$ is determined given $\Gamma$ and $s_c$. Recovering
the $s_c\to 0$ limit is subtle, since it also implies low mass $M$.
To summarize: for EOS MR, we use Eq. (\ref{eddington}) to solve for
$s_c$ given $M$ in terms of $M_{scale} \equiv  (m_p/2 \mu)^2M_\odot$.

In this paper we will adopt the point of view that $k$ is not known 
{\sl a priori} and we will allow scaling of the polytropic solution to
arbitrary $M$ and $R$ in cases with no radiation pressure. 
In cases with radiation pressure, although $M$ is determined from
Eq. (\ref{eddington}), some scaling remains since Eq. (\ref{mreqn})
relates $R$ and $k$, given $M$ and $\Gamma$, but does not determine
either one separately.

\subsection{Convective Stability of Polytropes}
\begin{figure}
\centering
\includegraphics[width=80mm]{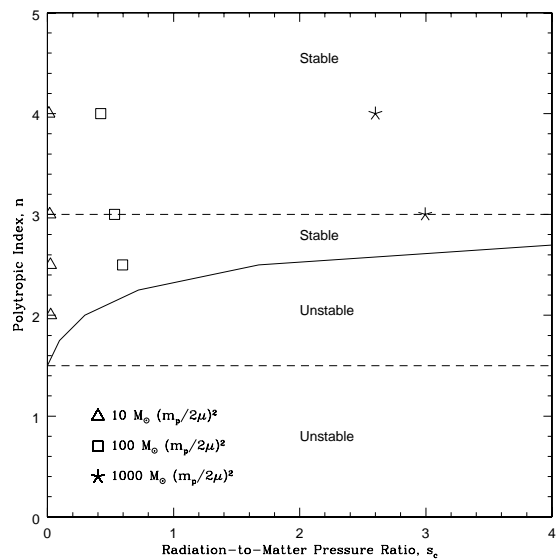}
\caption{\label{fig:stabstars} A plot summarizing the results of the stability analysis described in \S 4.2. 
Convectively stable and unstable regions of the $n$ - $s_c$ parameter space are labeled, 
and the models we have selected from this parameter space are marked.
}
\end{figure}

If we begin with the First Law of Thermodynamics,
\be
T dS = dE + PdV,
\ee
and use EOS MR, together with some of the relations derived in Appendix B, we can put it
in the form
\be
TdS=f(s,n)dP/\rho .
\ee
Since $dP/dr < 0$, local convective stability requires $dS / dP < 0$. 
The condition is
\be
n > n_{crit} = \frac{3(8s_r+1)(s_r+1)}{8s_r^2+13s_r+2},
\ee
where we have used $s_r = s(r) = P_{rad}(r) / P_{matter}(r)$ to distinguish between this function of $r$ and the constant parameter $s_c$.
For stability, we require that the local condition be satisfied throughout the model. The variation of $s_r$ depends upon $n$: 
\be
s_r(1+s_r)^4\propto{P^3\over\rho^4}\propto P^{{3-n\over 1+n}}.
\ee
There are three characteristic cases
\begin{enumerate}
\item For $n<3$, the largest value of $s_r$ is $s_c = s(r=0)$, so the largest
value of $n_{crit}$ is $n_{crit}(s_c)$ and therefore if
$n>n_{crit}(s_c)$, the stellar model is stable. 
\begin{itemize}
\item For all $0 \leq s_r \leq \infty$, Eq. (5) implies $n_{crit}(s_r)\geq 3/2$, so no polytrope with $n<3/2$ can be stable.
\item For $3/2<n<3$, there is a maximum value of $s_c$ for which the
stellar model is stable, determined from $n=n_{crit}(s_c)$, which rises from
$s_c=0$ at $n=3/2$ to $s_c\to\infty$ for $n=3$. For instance, 
at $n=2.0$, $s_c^{MAX} = 0.3$; for $n=2.5$, $s_c^{MAX} = 1.674$ (see Tab.~\ref{tab:starmasses}
and Fig.~\ref{fig:stabstars}).
\end{itemize}
\item For $n=3$, $s_r$ is constant. Since $n_{crit}(s_r)<3$ for any
finite $s_r$, all models are stable.
\item For $n>3$, $s_r \to \infty$ at the polytrope's edge (where $P \to 0$), so for any choice of $s_c$ the largest value of $n_{crit}(s_r)\equiv 3$. All models are stable.
\end{enumerate}
It is important to emphasize that the convective-stability condition described here is not an
absolute stability criterion. Stellar models outside of this mass range can certainly exist; they
will simply have convection zones. The unstable region in Fig. \ref{tab:starmasses}
arises because of our insistence on
an exact polytropic pressure-density relation ($P = \kappa \rho^{\Gamma}$) and our use of
EOS MR.  We restrict our studies to stable models.  These already span a wide
range of density profiles -- which is the basic physical property that governs shock propagation --
from the highly centrally-condensed $n=4$ polytropes to the diffuse $n=2$ models. An unstable
(polytropic) model would represent an unphysical and unnatural initial state.
Actual convective stars involve physics beyond the scope
of our simple, idealized, one-dimensional approach.

\begin{table}
\caption{\label{tab:starmasses} Summary of the parameters describing the models we have studied, where $\mdim = M_{star}/M_{scale}.$}
\begin{center}
\begin{tabular}{|c|c|c}\hline
$n$ & $\mdim $ & $s_c$\\\hline
2.0 & 10 & 0.0256\\
2.5 & `` & 0.0227\\
3.0 & `` & 0.0181 \\
4.0 & `` & 0.0117\\
2.5 & 100 & 0.5965\\
3.0 &  `` & 0.5515 \\
4.0 &  `` & 0.4219\\
3.0 & 1000 & 2.995\\
4.0 &   `` & 2.600 \\ \hline
\end{tabular}
\end{center}
\end{table}

\subsection{Description of Analysis}

\begin{figure}
\centering
\includegraphics[width=80mm]{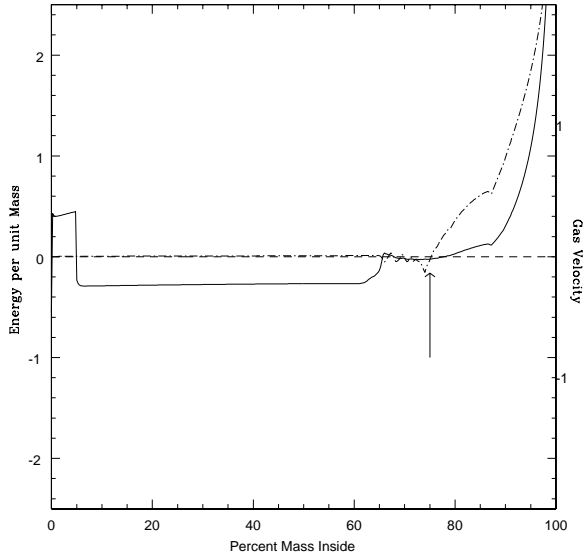}
\caption{\label{fig:echeck}The asymptotic dimensionless local 
energy per unit mass ($U / (GM_{star}/R_{star})$) (solid line) for $n=3$
polytrope (no radiation pressure) and $E_{\rm blast} = E_{\rm bind}$ after the
central core has reattained hydrostatic equilibrium.  The dash-dotted line gives the 
dimensionless local velocity ($v/(GM_{star} / R_{star})^{1/2}$). 
The dashed line is a reference line for zero energy and velocity.}
\end{figure}

The chief way in which we shall summarize the results of an explosion
is in terms of the mass ejected as a function of explosion energy. We
begin by discussing how we extracted the ejected mass from the numerical
simulations. For EOS M (no
radiation pressure), the code was run until the remnant core had
become stationary and had nearly reassumed hydrostatic equilibrium, \emph{i.e.}, it had local gas velocities near zero ($<10^{-7} R_{stellar} / {\rm dynamical \; time}$) and satisfied the virial theorem.
The mass loss was determined by finding the location in the Lagrangian
grid of the outermost outgoing zone for which the local energy per unit mass (sum of kinetic, thermal and
gravitational contributions) changed from negative to positive (see arrow in Fig. \ref{fig:echeck}).
A graph of the local energy density for a typical stellar model after an explosion is included
as Fig.~\ref{fig:echeck}.

A drawback of this method is apparent in Fig. \ref{fig:echeck}.  Though there are
distinct portions of the stellar model that can definitely be said to be either
remnant or ejecta, there is also a small region with nearly zero
energy, resembling an atmosphere around the remnant. These atmospheres
did not appear in all explosions -- typically, they occurred when
$E_{\rm blast} \sim$ 0.7 - 1.0 $E_{\rm bind}$. In some cases, it was adequate
simply to run these models longer, with a clear bifurcation point eventually 
emerging. 

\begin{figure}
\centering
\includegraphics[width=80mm]{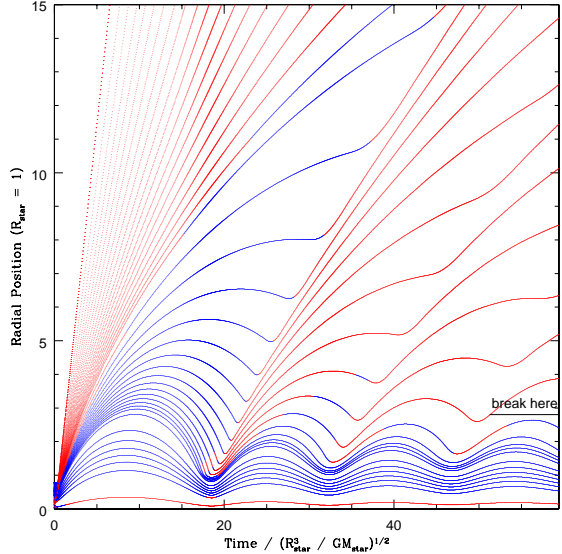}
\caption{\label{fig:typrad}A typical graph of the motion of grid zones in time for an
exploding stellar model, in this case an $\mdim = 100$
 polytrope of index $n=3$ with an explosion energy of $90$ per cent of the stellar model's binding energy.
 Each line tracks a representative mass shell.  The Lagrangian
mass intervals vary: lines in the ejected region and outer
parts of the remnant -- which represent escaping mass (large radii)
and the uppermost parts of the cooling, bouncing remnant (the blue lines)
-- are at intervals of $0.5-1$ per cent of the total mass. In the inner
part of the remnant, each line represents approximately $10$ per cent of the
total mass. Where lines are red, the local energy density is positive;
where blue, negative. The bifurcation point separating the remnant
 from the ejecta is marked. Radial distances are given in units of the
  initial stellar radius; times are given in dimensionless units defined by 
$t / (R^3_{star} / GM_{star})^{1/2}$.
}
\end{figure}

When we began to us EOS MR, however, our results remained ambiguous 
even after long integration times.
Thus, we moved to a more detailed procedure for deciding which
mass shells were ejecta and which composed the remnant (see Fig. \ref{fig:typrad}). We stored the
location and local energy density of each grid zone throughout the run. 
We then plotted the location of each mass element as a function of
time, using the sign of the local energy density to color code the
lines.  During the atmospheric motions some layers do work on other
layers; the color coding shows changes from bound to unbound (and
vice-versa).  These plots proved to be helpful, illuminating the
transient identities of bound atmospheres, marginally bound gas, and
low energy ejecta. We adopted the following criterion for ending the
calculation: when all outer shells had positive energy
density and the number of intermediate shells with local energy
density still changing sign was small -- less than a couple of percent of
the total mass. An example is shown in Fig.~\ref{fig:typrad}, where the apparent
bifurcation point between bound and unbound material is marked on the far right.
Note the diminishing amount of mass ejected with each stellar
oscillation. We are confident of this prediction because it was
borne out in all cases where the code was run much longer, and, hence,
closer to the point of the remnant's return to hydrodynamic
stability. 

These plots illustrate two distinct ways in which shells
are ejected in successful explosions that are not yet large enough
to destroy completely the stellar model: 1) Some
shells are lost in the initial shock wave; these gain
tremendous kinetic energies. The amount of mass lost in this way is typically $\le 12$ per cent.
2) The rest of the ejecta are expelled by the ringing down of the remnant.

We next investigated the extreme limits: total disruption
explosions and failed explosions (no ejected mass).  Total disruptions
were relatively easy to recognize: all grid zones acquired
positive energy in the first pass of the shock wave from the
explosion. For explosion energies near
the threshold for total stellar disruption, however, we found that there was
typically a range of energies where a large portion of the stellar envelope 
was ejected, while leaving an extended remnant undergoing
very slow, long-scale oscillations. The precise mass loss in most of these cases was 
impossible to determine, though it was always $\sim 50$ per cent.
Artificial viscosity eventually 
damps these oscillations,  but it takes a long time to do so.
Total disruption happens abruptly, with every stellar model studied going from the $\sim 50$ per cent
mass loss oscillatory state to 100 per cent mass loss with only a small increase in explosion
energy. We were able to pin down the width of the transition from
remnant to total disruption as function of explosion energy to $\sim
5-10$ per cent in the stellar model's binding energy.

\begin{figure}
\centering
\includegraphics[width=80mm]{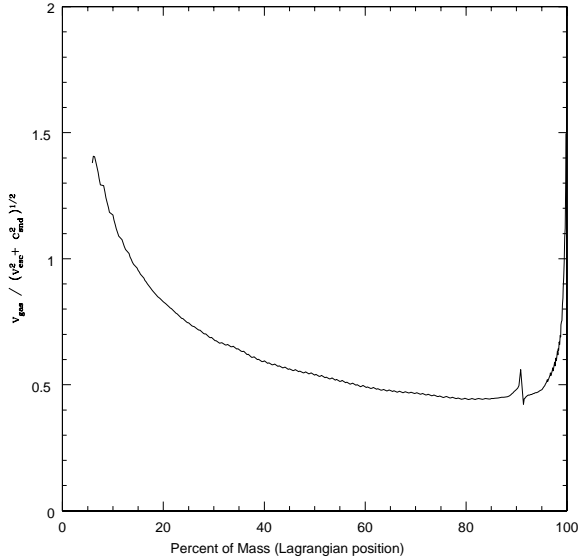}
\caption{\label{fig:sound}A plot of the $v_{gas} / \sqrt{v_{esc}^2 +
c_{snd}^2}$, in this case for a stellar model with $n=1.5$ , $\mdim =0.55$, and an
explosion energy equal to $15$ per cent of the stellar model's binding energy. The
choppiness in the plot is due to non-physical effects in the
determination of the exact location of the shock. Note the deceleration through
the bulk of the stellar model, with only the very outermost shells reaching
escape velocity as the shock accelerates in the falling density
profile near the edge.}
\end{figure}

The failed explosion regime was computationally easier to study.
Failed explosions produced no unbound shells.
The results can be understood in terms of the varying shock speed.
Strong shocks slowed as they plowed
through the dense core of the stellar model, then accelerated when they reached
the steep density gradient of the outer regions of the polytropes. In failed explosions the
shock velocity fell below the sound speed in the middle region and/or
failed to accelerate up to the local escape speed in the outer
region. A plot of the process is included in Fig.\ref{fig:sound}. In the figure,
we plot $v_{shock} / \sqrt{v_{esc}^2 + c_{snd}^2}$ -- 
where $v_{esc}$ is the escape velocity for the initial stellar model 
and  $c_{snd}$ is the local sound speed -- illustrating
the falling shock Mach number in the core and the reacceleration
to velocities allowing escape by the large, negative, outward-going
density gradient. The physics of this variable shock velocity, including
spherical curvature effects and photon loss, are discussed 
in detail in \cite{matz99}.

\subsection{Explosions without radiation pressure}

For the first round of explosions, we used EOS M (no radiation
pressure). For these calculations, the solution's independent parameters are
$E_{\rm blast}/E_{\rm bind}$ and the stellar model's polytropic index $n$.

\begin{figure}
\centering
\includegraphics[width=80mm]{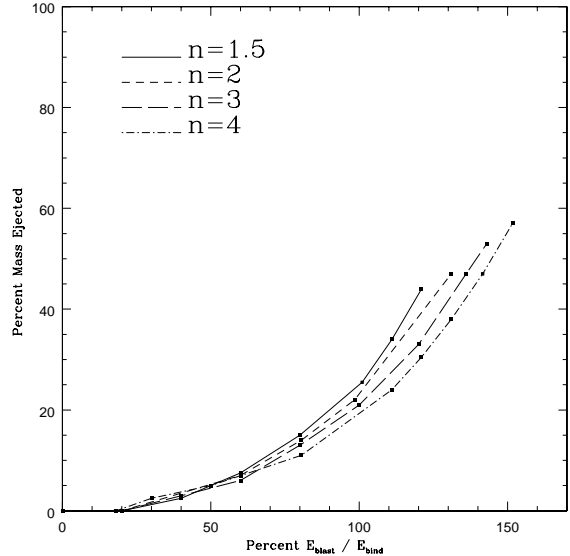}
\caption{\label{fig:compare}This figure summarizes the mass loss percentages resulting from explosions in polytropes  of four different indices without radiation.}
\end{figure}

\begin{figure}
\centering
\includegraphics[width=80mm]{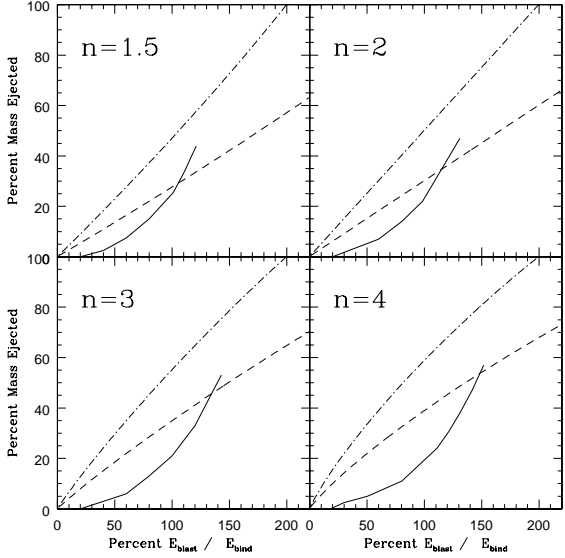}
\caption{\label{fig:sofour}Mass loss curves are compared with two simple assumptions relating explosion energy and mass loss. The dashed curve represents the most efficient possible application of the explosion energy to mass loss. The dash-dotted curve represents the more physically reasonable assumption that 50 per cent of the energy goes into unbinding part of the stellar model, and 50 per cent goes into both heating the remnant and to net kinetic energy for the ejecta.}
\end{figure}

We have included two figures summarizing the mass loss results. In
the first, Fig.~\ref{fig:compare}, the explosions are compared with each other,
showing great similarity among the models. In Fig.~\ref{fig:sofour}, we have
separated each polytrope into its own window to compare its mass loss
curve to a couple of ``rules of thumb'' 
\citep{fry01}. The first line, the dash-dotted curve, is the
simplest such rule.  It represents the mass loss if
100 per cent of the explosion energy were distributed in such a way as to
eject as many of the outer shells as possible,
while leaving untouched those parts of the stellar model which remain bound.
This is, of course, physically impossible, but it does
provide an upper bound on mass loss. The dashed curve represents the
actual choice made by \citet{fry01}, which essentially
splits the explosion energy budget in two, giving 50 per cent to unbinding
the stellar model directly, and 50 per cent to heating the remnant and to
accelerating the ejecta. 
This version gives results that are much closer to
our numerical calculation, but {\it overestimates mass loss in low energy
explosions} and also {\it overestimates the amount of energy required to
unbind the stellar model completely}.

\subsection{Explosions with radiation pressure}

\begin{figure}
\centering
\includegraphics[width=80mm]{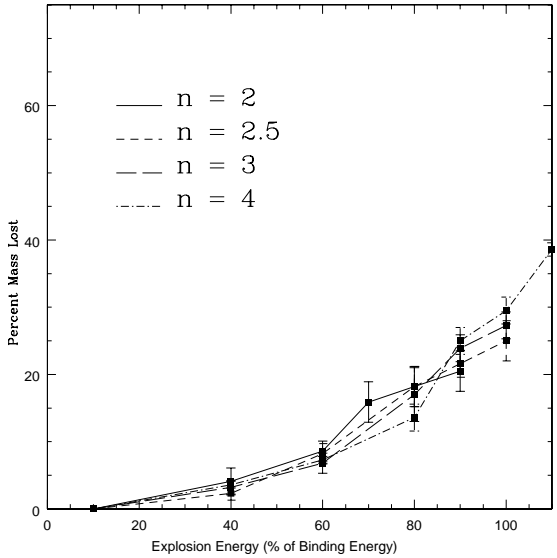}
\caption{\label{fig:tenM}This figure summarizes the mass loss percentages for explosions in $\mdim = 10$ stellar models. Error bars represent the uncertainty in the determination of the remnant - ejecta bifurcation point.}
\end{figure}

\begin{figure}
\centering
\includegraphics[width=80mm]{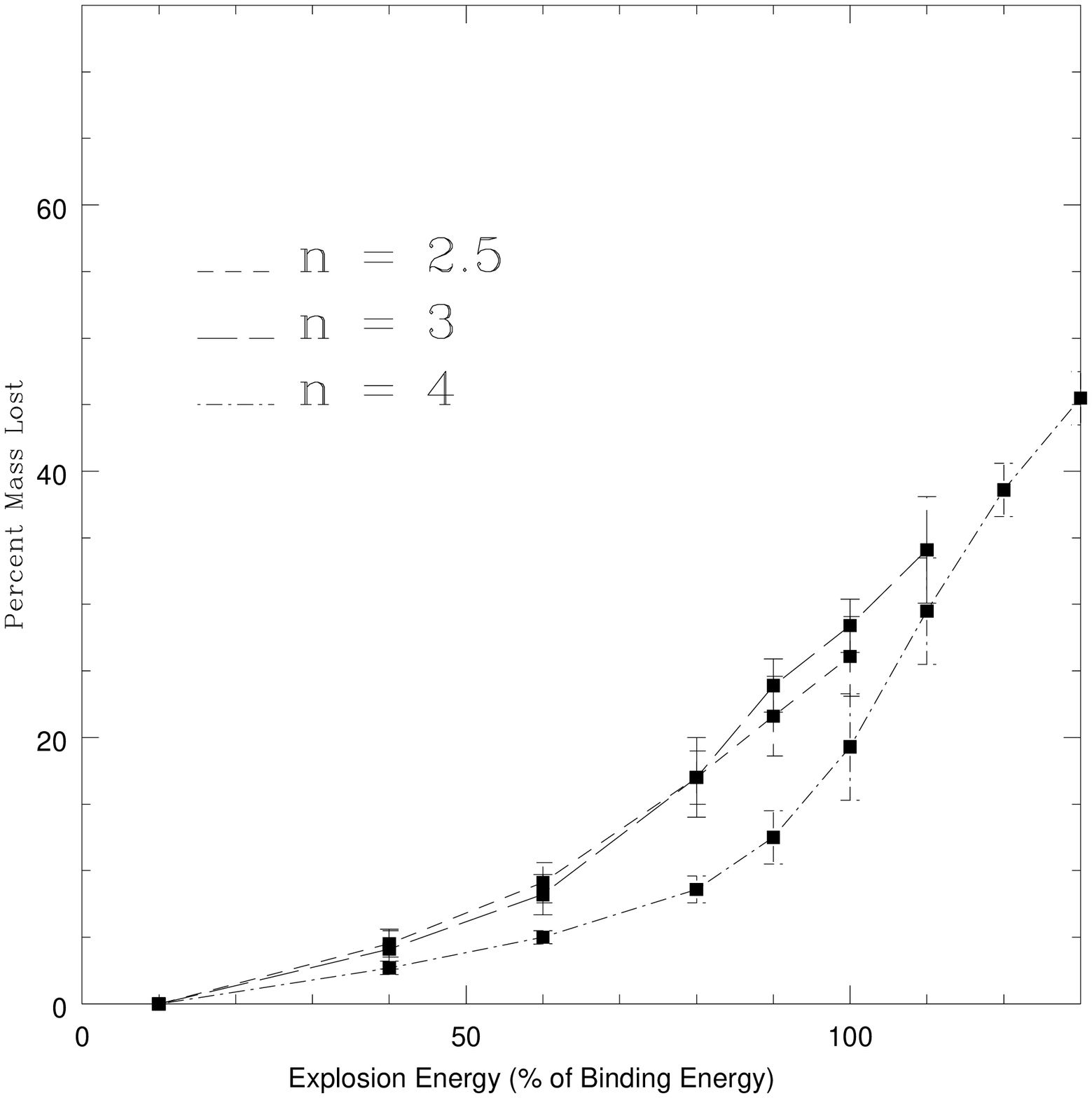}
\caption{\label{fig:hundM}This figure summarizes the mass loss percentages for explosions in $\mdim =100$ stellar models. Error bars represent the uncertainty in the determination of the remnant - ejecta bifurcation point.}
\end{figure}

\begin{figure}
\centering
\includegraphics[width=80mm]{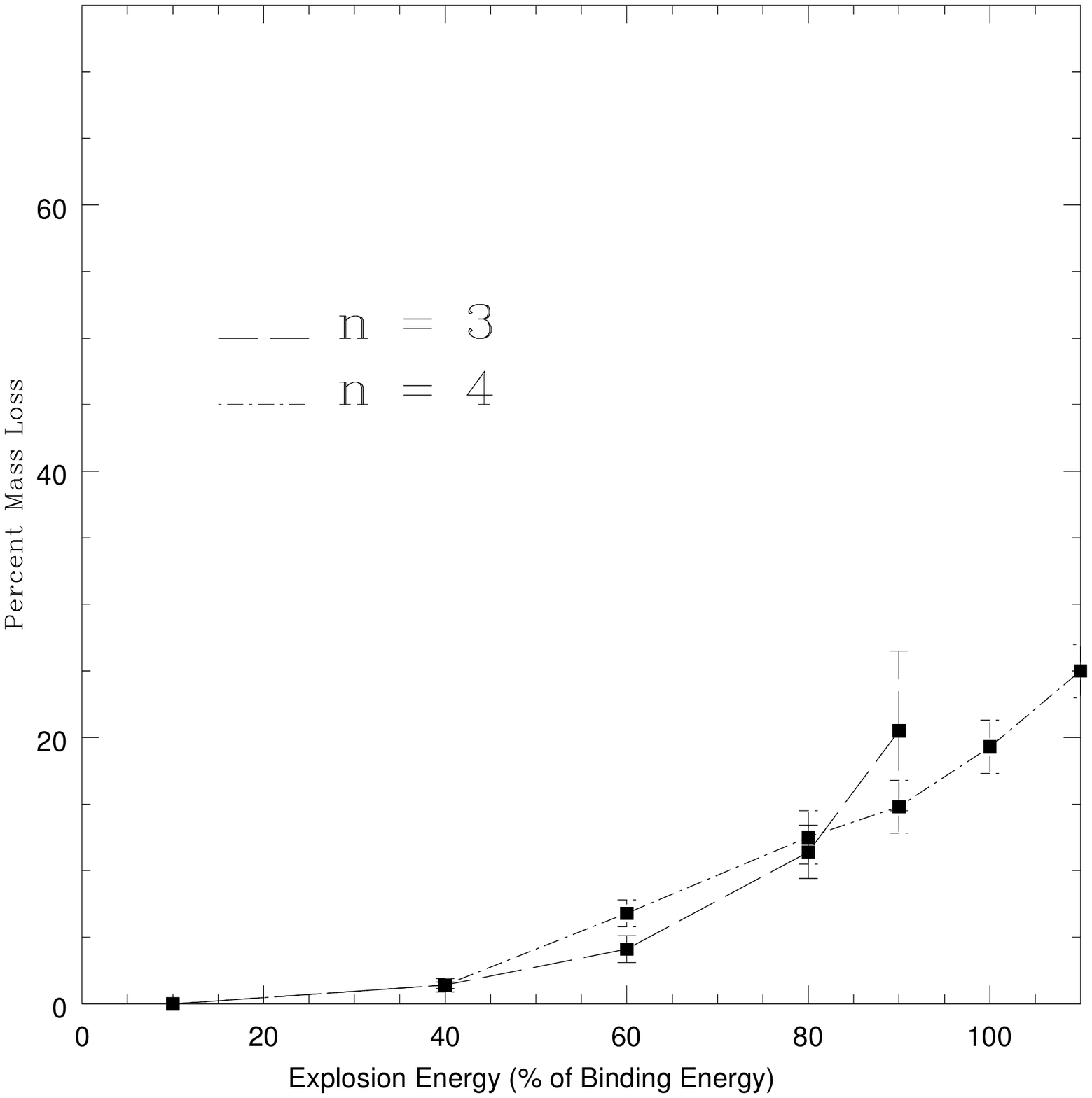}
\caption{\label{fig:thouM}This figure summarizes the mass loss percentages for explosions in $\mdim = 1000$ stellar models.  Error bars represent the uncertainty in the determination of the remnant - ejecta bifurcation point.}
\end{figure}

For the second round of explosions, we used the hydrodynamics code
with EOS MR (matter and radiation pressure). The parameter
space now included three variables: explosion energy, polytropic
index, and $M_{star}$. We chose three stellar masses,
$\mdim = 10,\,100,$ and $1000$, and then surveyed the range of polytropic 
indices for convectively stable models. For each stellar model,
we varied explosion energy from cases of failed explosions to total stellar disruption.
The results of these explosions are summarized in Figs.~\ref{fig:tenM}, \ref{fig:hundM}, and  \ref{fig:thouM}. 
Because of the difficulty of precisely determining the line of bifurcation between remnant
and ejecta even in plots like Fig. \ref{fig:typrad}, error bars have been included. They represent
 the range within the stellar model where the bifurcation point may occur.
 This range of uncertainty was determined by finding the region of the 
stellar model containing either outgoing zones with negative local energy density, located
at several stellar radii, or infalling zones with positive local energy density. 
We considered these to have ambiguous fates.
 The number of these zones is always small, comprising at the 
most a couple of percent of the stellar mass. Bold dots are placed at the midpoint 
of this range.

\begin{figure}
\centering
\includegraphics[width=80mm]{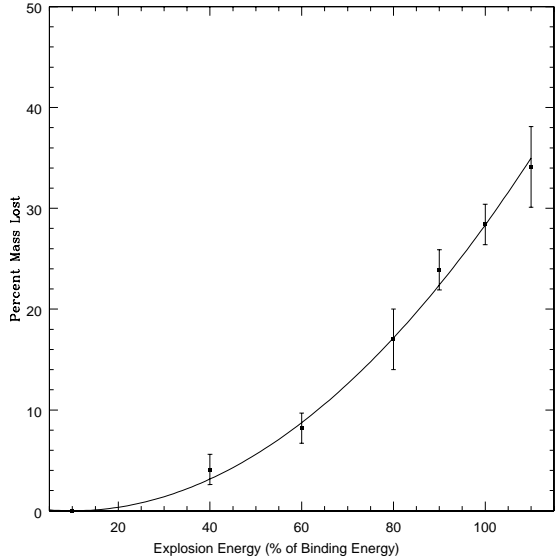}
\caption{\label{fig:fitform} A figure comparing the results of the fitting formula, Eqn.~\ref{fitdata}, 
to the data, in this case for an $n=3$, $\mdim = 100$ stellar model.}
\end{figure}

The mass-loss curves are remarkably uniform, especially given 
the wide variation in binding energy among the stellar models that we studied
(more details on binding energy are given in Appendix B). 
The uniformity in the shapes of the mass-loss curves found allows
them to be described accurately by a fitting formula. The form that best
fits the data is
\begin{equation}
100 \times \frac{M_{\rm lost}}{M_{star}} = \left \{ \begin{array}{ll} 
0 & e < e_o \\
A(e_{\rm blast}-e_o)^2 & e_o < e < e_f \\
100 & e_f < e
\end{array} \right.
\label{fitdata}
\end{equation}
where $e_{blast}$ ($e_o$, $e_f$) is blast energy (minimum
blast energy to cause mass loss, maximum blast energy to leave a bound
core) measured as a percent of binding energy 
(\emph{i.e.} $e_{\rm blast} = 100 \times E_{\rm blast} / E_{\rm bind}$, etc.) and $A$ is the fitting
parameter. A sample comparison between the fits and the numerical
data is shown graphically in Fig.~\ref{fig:fitform}; the parameters describing each
model's fit are contained in Table~\ref{tab:massloss}.

Notice that in all cases, total disruption occurs near or slightly above the
original stellar binding energy, \emph{i.e.} at $E_{\rm blast} = (e_f/100) E_{\rm bind} \ga E_{\rm bind}$.
The transition to total disruption is also very abrupt. Just below $e_f$, the amount of mass lost 
is below 50 per cent in all cases.

\begin{table}
\caption{\label{tab:massloss}Summary of fitting parameters for various explosion scenarios. $e_o = 100 \times E_o / E_{\rm bind}$, $e_f = 100 \times E_f / E_{\rm bind}$}
\begin{center}
\begin{tabular}{|c|c|ccc|}\hline
$n$ & $\mdim$ & $A$($\times 10^{-3}$) & $e_o$ & $e_f$ \\\hline
2.0 & 10 & 3.57 & 10 & 110\\
2.5 & `` & 3.37 & 11 & 120\\
3.0 & `` & 3.47 & 10 & 120 \\
4.0 & `` & 3.67 & 10 & 130\\
2.5 & 100 & 2.89 & 5 & 120\\
3.0 &  `` & 3.50 & 10 & 130 \\
4.0 &  `` & 2.95 & 11 & 140\\
3.0 & 1000 & 3.74 & 20 & 110\\
4.0 &   `` & 2.44 & 10 & 130\\ \hline
\end{tabular}
\end{center}
\end{table}

\section{Sensitivity to Inner Boundary Conditions}

Our previous results show that mass loss occurs
in two distinct phases when the stellar model is not
completely unbound. The initial
shock wave drives off some mass immediately while the
ringing down of the remnant expels loosely
bound outer shells over a period of several dynamical times.
None of our models included a compact object at the centre
and one naturally wonders whether the later mass
loss might be sensitive to the treatment of the centre of the stellar model.

 
To test the effect of the inner regions on the mass
loss, we considered two different ways of modifying the inner boundary condition.
In one, we placed a hard, reflecting sphere at a fixed, very small
spatial radius; in the other, we fixed a perfectly absorbing
boundary at a given radius. 
We ran several cases: polytropes of indices
$n$ between 1.5 and 4, with a variety of values of $\mdim$.
For the reflecting sphere, the size was set to that of the innermost zone;
for the absorbing sphere, the absorbing boundary
varied between 2 and 30 percent of the mass in Lagrangian
coordinates.

The mass loss computed with these altered inner boundary conditions differed little
from what was found in our earlier survey.
The changes to the mass loss were unnoticeable even 
when quite a large inner region -- as much 
as 15 - 20 per cent, in mass -- of the stellar model was allowed to become 
pressure-free and perfectly absorbing. This insensitivity is linked to the fact that
at least 50 per cent of the star remains bound if the star survives the explosion.
In those cases, on the one hand, the trading of energy among the outer shells 
determines the amount of mass lost; the inner half of 
the stellar model, whose dynamics are sensitive to the treatment of the inner zones, 
is never in danger of being lost.  In stellar models that are totally 
dispersed, on the other hand, the explosion is so violent that differences in 
the inner boundary conditions have little effect, since no portion of the 
stellar model ever falls back to experience them.

To gain a better qualitative understanding of why
the mass loss results are so insensitive to the inner boundary condition, we examined
more closely the slowly-varying disturbances propagating in our remnant.
The energy density in a small-wavelength oscillatory mode
is proportional to local density times the square of local
gas displacement (\emph{e.g.} \citealt{jorgen}). We computed this combination 
-- as well as local velocity squared times
local density -- as a function of radius at various times in our models,
using each Lagrangian zone's displacement from its final, at-rest location.
We found that both quantities always go to zero at the remnant core,
implying that there is little energy flux into or out of this region.
 Therefore, information about the altered inner
boundary conditions is not effectively transmitted to the oscillating
outer mass shells.  The inner
boundary condition plays no obvious role in the late-time mass loss
so long as there is a sufficient amount of buffering gas between the
inner core and the outside of the star. In our numerical tests, we found that it was necessary 
to alter almost the entire inner quarter of the stellar model's mass before
the mass loss was significantly affected. 

\section{Conclusion}

Here, we have confined ourselves to a rather idealized problem, 
partially disruptive explosions of simplified stellar models
whose density profiles are solutions to the Lane-Emden equation,
but with varying ratios of radiation to matter pressure. 
In this way, we have been able to survey models in a well-defined
parameter space fairly extensively, trading the concreteness of 
real stellar models for the flexibility of parametrized ideal models. Specifically,  we have
determined the fraction of the original stellar mass ejected as
a function of explosion energy
in polytropes of index $n = 1.5$, $2$, $3$,
and $4$ in calculations without radiation pressure; we also explored
the mass loss fractions for stellar models of $\mdim = 10,\, 100,$ and $1000$ over a range of 
$n$ from 2.0 to 4.0. Our
results suggest that the mass loss is remarkably uniform, even
among widely varying stellar density profiles and over an
enormous range in stellar masses.
We have provided a simple, parametrized formula for the
fractional mass loss as a function of explosion energy for
a range of values of $n$ and the ratio of radiation to matter
pressures; see Eq. (\ref{fitdata}) and Table~\ref{tab:massloss}.

One striking feature of all the models we tested was that the mass loss 
fraction as a function of explosion energy appears to be discontinuous 
at around 50 per cent mass loss, with a small (few per cent) difference in 
explosion energy separating stellar models which lose half their mass from 
totally disrupted stellar models. Because our simulations did not include
formation of a compact remnant at the centre, this result cannot
be taken as a concrete demonstration that the observation of a black
hole of mass $M$ demands a progenitor whose mass was
less than about twice as large.

The abrupt transition between moderate (\emph{i.e.} $\la
50$ per cent) and total disruption found here for wide classes
of initial models is also seen in modelling of explosions in
sets of specific progenitors (\emph{e.g.} \citealt{ww95},
Table 3; \citealt{mac01}, Table 1). Thus,
we conjecture that even when a compact central remnant is
included, the results divide into two separate cases depending
on whether the explosion energy is above or below, approximately,
the critical value found here for complete disruption. It is significant that this
 bifurcation effect occurs, as it does in our simulation, at the most basic, 
 hydrodynamic level, implying that its appearance in more sophisticated 
 models rests chiefly on these underlying physics. For explosion
energies below this critical value, there is a sharp transition
between modest (\emph{i.e.} $\la 50$ per cent) mass loss and total disruption
apart from the compact remnant. For explosion energies above the
cutoff, either a black hole or neutron star may form. However, in this case, we expect much smaller
fallback masses, generally only a few tenths of $M_{\sun}$ or less,
primarily caused by reverse shock propagation through the core, and
the consequent deceleration of a small amount of outgoing matter
(\emph{e.g.} \citealt{woos88, chev89}).

We should emphasize that our conjectures here may be significantly 
modified by the inclusion of more realistic physics. Spherically-symmetric polytropic stellar models
are not real stars; neither are all the complexities of stellar hydrodynamics 
captured by simple equations of state.
One important component of realistic stars that our models lack is the sequence of density 
discontinuities that emerge in evolved stars, which we would expect to cause 
reflected reverse shocks that could change the mass ejection dynamics considerably.
Furthermore, in real supernovae there are other sources of energy besides the initial 
explosion -- such as decaying radioactive nuclei, central pulsars or jets
-- which continue to add energy
to the supernova system at late times, possibly giving the boost needed to eject
lightly-bound gas at the surface of the remnant.  Our imposition of 
spherical symmetry is also unrealistic: real explosion are likely to be
anisotropic, and stellar rotation, convection, and magnetic fields are expected
to have effects that a one-dimensional calculation cannot hope to model. 
Despite these caveats, we believe our chief results to be robust. 

Supernovae frequently occur in binary systems. In such systems, when there are very energetic explosions and little mass fallback, we
expect little mass contamination of the atmosphere of the binary
companion.  The outgoing regions of the progenitor intercepted by the
companion are not captured; indeed the outer layers of the companion
are stripped and ablated by the ejecta. On the other hand, when the
explosion is weak and substantial mass fallback occurs, progenitor
material may fall back onto the companion, polluting its
atmosphere. In the latter cases, we would then infer that a remnant of
mass $M$ was most likely derived from a progenitor with mass less than
$\simeq 2M$. Thus, in systems like Nova Scorpii that show evidence for
black hole formation in a supernova (\emph{e.g.} \citealt{isr99}), we
conjecture that mass of the pre-explosion star was, in fact, less than
twice the present mass inferred for the black hole remnant (which also
has accreted matter since forming, presumably). This may have
implications for the dynamics of such systems (\emph{e.g.} \citealt{mir02}).  We caution, though, that our results may be altered somewhat
in more refined models. Further studies are underway to include a
compact central remnant, density jumps (expected as a consequence of
compositional inhomogeneity), rotation and explosion
asymmetries. These new calculations will continue, in the same spirit
as those reported here, to employ the simplest explosion models needed
to reveal the underlying physical consequences of the various
refinements, and to allow a survey of the hydrodynamics of a
large range of explosion models.

\section*{Acknowledgments}
This research was supported in part by NASA-ATP grant
NAG5-8356. M.W. is supported by an NSF Graduate Fellowship.
I.W. acknowledges the hospitality of KITP, which is supported by NSF
grant PHY99-07949, where part of this research was carried out, as well
as support from NSF Grant AST-0307273 and from IGPP at LANL.

\appendix
\section{Difference Equations}
The Lax-Wendroff difference equations for the equations of hydrodynamics in one dimension with spherical symmetry are as follows. Note that the pressure in the equation for advancing energy must be solved for using the equation of state to make this set of difference equations explicit rather than implicit.  
In the difference equations, n represents time steps, while j represents spatial steps. The equations are non-dimensionalized simply, with each variable scaled to order unity for the initial conditions in all calculations we have done. The one remaining constant, $\rho_o$, with units of density, sets the overall scale of the system studied. The variable $R$ records the position of each shell. Comparing each shell's current position, $R$, with $r$, a static, reference grid, allows the gas's local density to be calculated. The remaining variables are interdependent. 
The equation for moving grid zones is:
\begin{equation}
\frac {R_j^{n+1} - R^n_j} {\Delta t}   =  u_j^{n+1}.
\end{equation}
The conservation of momentum equation is:
\begin{equation}
\frac {u_j^{n+1} - u_j^n} {\Delta t}   =  - \frac 1 {\rho_o} \frac { (\delta p)^n_j} {\Delta r} \left ( \frac {R^n_j} {r_j} \right )^2.
\end{equation}
The conservation of mass equation is:
\begin{equation}
\rho_{j+1/2}^{n+1}   =  \rho_o \frac { (r_{j+1} )^3 - (r_j)^3} { (R_{j+1}^{n+1})^3 - (R_j^{n+1})^3 } .
\end{equation}
The First Law of Thermodynamics is:
\begin{eqnarray}
U_{j + 1/2}^{n+1}  & = &  U_{j+1/2}^n - \left ( \frac {p_{j+1/2}^{n+1} +p_{j+1/2}^n} 2 \right ) \times \nonumber \\
& & \left ( \frac 1 {\rho_{j+1/2}^{n+1}} - \frac 1 {\rho_{j+1/2}^n} \right ).
\end{eqnarray}
Where $U =$ internal energy / mass.  The acceleration of the innermost shell is determined by treating its volume as filled with a gas of uniform pressure so that the shell's equation of motion is:
\begin{eqnarray}
m_{inner} \frac {\partial v} {\partial t} & = & 4 \pi (p_{inner} - p_{outer}) \Rightarrow \nonumber \\
 u_0^{n+1} &= & u_0^{n} + 4 \pi \Delta t (p_{inner} - p_{outer}).
\end{eqnarray}
The pressure within the inner sphere varies adiabatically as the shell moves, \emph{i.e.},
\begin{equation}
p_{inner} (t) = p_o \left ( \frac {V_o} {V(t)} \right )^{\gamma}. 
\end{equation}

These equations are completed by some equation of state,
\begin{equation}
p_{j+1/2}^{n+1}  =  f ( U_{j+1/2}^{n+1}, \rho_{j+1/2}^{n+1} ).
\end{equation}
If this equation of state can be algebraically solved, the full set of equations is explicit; if it cannot be solved, then an implicit step and numerical root-finding procedure is required to advance the grid. 
The advancement of the grid proceeds as follows. 1) Using the conservation of momentum, the new gas velocities are set throughout the system. 2) Boundary conditions are applied. 3) The shell position, $R$, is advanced according to the new gas velocities. 4) R is then used to set the density throughout the system. 5) Two possibilities: if the equation of state is explicitly soluble, then the internal energy of the gas is determined. If not, then the pressure and energy equations must be stated in terms of the temperature and then solved, together with the First Law, numerically -- three equations for three variables, p, U, and T. 
The above prescription must be modified slightly to accommodate shock fitting. To this end, we introduce an artificial viscous pressure, q, given by the differenced form,
\begin{equation}
q_{j+1/2}^n  =  \left \{ \begin{array}{ll}
\frac {2 a^2 [ (\delta u)_{j+1/2}^n ]^2} { 1/{\rho^n_{j+1/2} + 1/\rho_{j+1/2}^{n-1}} } & \textrm{if $(\delta u)_{j+1/2}^n < 0$ }\\ 0 & \textrm{if $(\delta u)_{j+1/2}^n \geq 0$} \end{array}\right. 
\end{equation}
Note the parameter, a, which controls how widely the viscous pressure spreads the shock. Optimal values are $1.5<a<2.0$, which spread the shock over 3-10 zones.
This artificial viscous pressure is added to the regular gas pressure in the above equations as follows:
In the conservation of momentum equation, 
\begin{equation}
(\delta p)^n_j \rightarrow (\delta p)^n_j + (\delta q)^n_j
\end{equation}
and in the energy conservation equation,
\begin{equation}
\frac {p_{j+1/2}^{n+1} +p_{j+1/2}^n} 2 \rightarrow \frac {p_{j+1/2}^{n+1} +p_{j+1/2}^n} 2 + q_{j+1/2}^{n+1}.
\end{equation}
When advancing the grid with artificial viscous pressure, the artificial viscosity term, q, is advanced before the energy equation, step 5 in the previous description.
\section{Setting Up Mass Shells in a Polytrope}

In the initial configuration whose mass and radius are $\mstar$
and $\rstar$, define a mass coordinate $\mhat=M/\mstar$
so that the shell is at radius $\rhat(\mhat)=R(M)/\rstar$. 

Let the pressure and density be $P(M)=\phat(\mhat)(G\mstar^2/\rstar^4)$
and $\rho(M)=\rhohat(\mhat)\mstar/\rstar^3$, respectively. At the centre
of the stellar model, we can find $\phat(0)=f_P(n)$ and $\rhohat(0)=f_\rho(n)$,
where $n$ is the polytropic index. At any other point in the stellar model, we
have $\rho(M)=\rho(0)\theta^n$ and $P(M)=P(0)\theta^{n+1}$. Thus,
we have
\be
\phat(\mhat)=f_P(n)[\theta(\mhat)]^{n+1}~~~~~
\rhohat(\mhat)=f_\rho(n)[\theta(\mhat)]^n~.
\ee
Usually, we specify the Lane-Emden function as a function of a dimensionless
radius. Getting $\rhat(\mhat)$ then requires a little bit of work. The
method is this: the mass inside a physical radius $R$ is
\be
M(R)=4\pi\int_0^R{dr~r^2~\rho(r)}=4\pi r_{\rm scale}^3\int_0^{x}
{dx~x^2~[\theta(x)]^n}~,
\ee
where $r_{\rm scale}$ is the radius scale. Divide
by the total mass to get the conversion equation
\be
\mhat={\int_0^x{dx~x^2~[\theta(x)]^n}\over\int_0^{x_0(n)}
{dx~x^2~[\theta(x)]^n}}={\int_0^{x_0(n)\rhat(\mhat)}
{dx~x^2~[\theta(x)]^n}\over\int_0^{x_0(n)}{dx~x^2~[\theta(x)]^n}}~,
\ee
where $\theta(x_0(n))=0$ and we have used the fact that the scaled
Lane-Emden radius variable $x=x_0(n)\rhat(\mhat)$. This equation can be inverted numerically
to get $\rhat(\mhat)$, and we can use the result to evaluate the
pressure and density:
\ba
\phat(\mhat)& =&f_P(n)[\theta(x_0(n)\rhat(\mhat))]^{n+1} \nonumber \\
\rhohat(\mhat)&=&f_\rho(n)[\theta(x_0(n)\rhat(\mhat))]^n \nonumber .
\ea

We are interested in polytropic models where the pressure is 
supplied by a mixture of radiation and a nonrelativistic gas.
The total pressure in physical units is then
\be
P={\rho kT\over\mu}+{1\over 3}aT^4=P_{\rm gas}+P_{\rm radiation}~,
\ee
where $\mu$ is the mass per nonrelativistic particle in the 
stellar model. We define s
\be
s={P_{\rm radiation}\over P_{\rm gas}}={aT^3\mu\over 3\rho k}~,
\ee
which varies throughout the stellar model. At any point in the stellar model, we
can use this to eliminate temperature in favor of $s$:
\ba
T=&\left({3\rho s k\over a\mu}\right)^{1/3}& 
\Rightarrow \nonumber  \\
 P=&{\rho kT(1+s)\over\mu}= &
\left({\rho k\over\mu}\right)^{4/3}
\left({3s\over a}\right)^{1/3}(1+s)~.\nonumber
\ea
From this it follows that
\baray
{P\over P(0)}&=&\left({\rho\over\rho(0)}\right)^{1+1/n}
=\left[\theta\left((x_0(n)\rhat(\mhat)\right)\right]^{n+1}\nonumber\\
&=&\left({\rho\over\rho(0)}\right)^{4/3}
\left({s\over s(0)}\right)^{1/3}\left({1+s\over 1+s(0)}\right)
\nonumber\\
&=&\left[\theta\left((x_0(n)\rhat(\mhat)\right)\right]^{4n/3}
\left({s\over s(0)}\right)^{1/3}
\left({1+s\over 1+s(0)}\right)\nonumber~,
\earay
We can use this to solve for $s(\mhat)$ via
$$
[s(\mhat)]^{1/3}[1+s(\mhat)]
=[s(0)]^{1/3}[1+s(0)]\left[\theta\left(x_0(n)\rhat(\mhat)\right)\right]
^{1-n/3}.
$$
The thermal or internal energy density inside the stellar model is, in physical units,
\ba
\rho U &= &{3\rho kT\over 2\mu}+aT^4 ~ = {3\rho kT\over 2\mu}(1+2s) \nonumber \\
&=&{3\over 2}\left({\rho k\over\mu}\right)^{4/3}\left({3s\over a}
\right)^{1/3}(1+2s)~;
\ea
compare this with the pressure
\be
P=\left({\rho k\over\mu}\right)^{4/3}\left({3s\over a}\right)^{1/3}
(1+s)~.
\ee
Define $\gamma(\mhat)$ by $P=(\gamma(\mhat)-1)\rho U$, to find that
\be
{\rho U\over P}=(\gamma(\mhat)-1)^{-1}={3\over 2}
\left[{1+2s(\mhat)\over 1+s(\mhat)}\right]~;
\ee
let $U=(G\mstar/\rstar) \uhat$ to find that
\be
\uhat(\mhat)={3f_P(n)\over 2f_\rho(n)}\theta\left(x_0(n)\rhat(\mhat)\right)
\left[{1+2s(\mhat)\over 1+s(\mhat)}\right]~.
\ee
This completes the setup of the initial conditions of the polytrope.

We will also need to have the total energy of the stellar model because we want
to choose the blast energy as a fraction of the binding energy.
The gravitational energy of a polytrope of index $n$ is
\be
E_{\rm grav}=-{3G\mstar^2\over (5-n)\rstar}~.
\ee
The internal energy of the stellar model is 
\baray
E_{\rm int}&=&4\pi\int_0^{\rstar}{dr~r^2~\rho(r)U(r)}
=\int_0^{\mstar}{dM~U(M)}\nonumber\\
&=&{3f_P(n)G\mstar^2\over 2f_\rho(n)
\rstar}\int_0^1{d\mhat~\theta(x_0(n)\rhat(\mhat))\left[{1+
2s(\mhat)\over 1+s(\mhat)}\right]}\nonumber.
\earay
This is enough to get the total energy, but we can be a slight
bit more elegant by using the virial theorem,
\ba
-E_{\rm grav}&=&3\int{dM~{P\over\rho}} \nonumber\\
&=&{3f_P(n)G\mstar^2\over
 f_\rho(n)\rstar}\int_0^1{d\mhat~\theta(x_0(n)\rhat(\mhat))}~,
\ea
from which we find that
\ba
E_{\rm tot}&=&-{3G\mstar^2\over 2(5-n)\rstar} \times \nonumber \\ 
&&\left[{\int_0^1{d\mhat~
\theta(x_0(n)\rhat(\mhat))[1+s(\mhat)]^{-1}}\over
\int_0^1{d\mhat~\theta(x_0(n)\rhat(\mhat))}}\right]~.
\ea
If we define $E_{\rm tot}=-\ehat_{\rm tot}[3G\mstar^2/2(5-n)\rstar]$
we see that
\baray
\label{etot}
\ehat_{\rm tot} & =& {\int_0^1{d\mhat~\theta(x_0(n)\rhat(\mhat))
[1+s(\mhat)]^{-1}}\over\int_0^1{d\mhat~\theta(x_0(n)\rhat(\mhat))}} \\ & = & {\int_0^1{d\mhat~\frac{\phat(\mhat)}{\rhohat(\mhat)}
[1+s(\mhat)]^{-1}}}\over\int_0^1{d\mhat~\frac{\phat(\mhat)}{\rhohat(\mhat)}}~.
\earay

\section{Dynamical Equations}

The equation of motion for a mass shell is
\ba
{\partial^2 R(M,t)\over\partial t^2}&=&-4\pi R^2(M,t){\partial P(M,t)
\over\partial M}- \nonumber \\
&& ~~~~~~~~{GM\over R^2(M,t)}+\avis(M,t)~,
\ea
where $\avis(M,t)$ is the viscous acceleration (which we include
using a prescribed artificial viscosity).
Introducing our nondimensional radius, pressure and mass implies
\ba
\rstar{\partial^2\rhat(\mhat,t)\over\partial t^2}
&= &{G\mstar\over\rstar^2}\left[-4\pi\rhat^2(\mhat,t){\partial\phat(\mhat,t)
\over\partial\mhat}- \right. \nonumber \\
&&~~~~~~\left. {\mhat\over\rhat^2(\mhat,t)}\right]+\avis(\mhat,t)~;
\ea
Define a dimensionless time by $t=(\rstar^3/G\mstar)^{1/2}\tau$; then
\ba
{\partial^2\rhat(\mhat,\tau)\over\partial\tau^2}&=&-4\pi\rhat^2(\mhat,\tau)
{\partial\phat(\mhat,\tau)\over\partial\mhat}\nonumber \\&&~~~~~~-{\mhat\over\rhat^2(\mhat,\tau)}
+\ahatvis(\mhat,\tau)~,
\ea
where the viscous acceleration is defined by $\avis(M,t)=\ahatvis(\mhat,\tau)
(G\mstar/\rstar^2)$.
Since we shall actually want equations that are first order in time,
we note that the radial velocity is
\ba
{\partial R(M,t)\over\partial t}&=&\left({G\mstar\over\rstar}\right)^{1/2}
{\partial\rhat(\mhat,\tau)\over\partial\tau}\nonumber \\
&=&\left({G\mstar\over\rstar}
\right)^{1/2}\vhat(\mhat,\tau)~,
\ea
and therefore
\baray
{\partial\rhat(\mhat,\tau)\over\partial\tau}&=&\vhat(\mhat,\tau)\nonumber\\
{\partial\vhat(\mhat,\tau)\over\partial\tau}&=&-4\pi\rhat^2(\mhat,\tau)
{\partial\phat(\mhat,\tau)\over\partial\mhat} \nonumber \\&&~~~~~~
-{\mhat\over\rhat^2(\mhat,\tau)}
+\ahatvis(\mhat,\tau)~.
\label{dynamics}
\earay
From the first law of thermodynamics, we get that
\be
{\partial U(M,t)\over\partial t}=\qvis(M,t)-P(M,t){\partial\over\partial t}
\left[{1\over\rho(M,t)}\right]~,
\ee
where $\qvis(M,t)$ is the viscous heating, which we take to be
$$
\qvis(M,t)  =  \left \{ \begin{array}{ll}
- a^2 \rho(M,t) \left ( \frac{\partial v}{\partial R} \right)^2 {\partial\over\partial t}
\left[{1\over\rho(M,t)}\right]  & \textrm{if $\left (\frac {\partial v}{\partial R} \right) < 0$ }\\ 0 & \textrm{if $\left ( \frac{\partial v}{\partial R} \right) \geq 0$} \end{array}\right.
$$
where $a$ is a constant with units of length. Introducing the same
nondimensional variables as in the polytrope setup we find that
\ba
{\partial\uhat(\mhat,\tau)\over\partial\tau}&=& - \left (a^2 \rhohat(\mhat, \tau) \left ( \frac{\partial \vhat}{\partial \rhat} \right)^2
+ \phat(\mhat,\tau) \right) \times \nonumber \\ && {\partial\over\partial\tau}
\left[{1\over\rhohat(\mhat,\tau)}\right]~,
\ea
where the viscous heating is defined by $\qvis=\qhatvis[(G\mstar)^{3/2}
/\rstar^{5/2}]$. We can replace the pressure by 
\be
P={2\rho U\over 3}\left({1+s\over 1+2s}\right)
~~\Rightarrow~~\phat={2\rhohat\uhat\over 3}\left({1+s\over 1+2s}\right)
\ee
to rewrite the first law in the form
\ba
{\partial\uhat(\mhat,\tau)\over\partial\tau} &= & 
\left ( \frac{a^2}{\rhohat(\mhat,\tau)} \left ( \frac{\partial \vhat}{\partial \rhat} \right)^2 + \right. \nonumber \\ 
&&\left. {2\uhat(\mhat,\tau)[1+s(\mhat,\tau)]\over
3\rhohat(\mhat,\tau)[1+2s(\mhat,\tau)]}
\right ) {\partial\rhohat(\mhat,\tau)\over\partial\tau}~.
\label{firstrho}
\ea
We can close the loop using $U\propto\rho^{1/3}s^{1/3}(1+2s)$ to
get
\be
{\uhat(\mhat,\tau)\over\uhat(\mhat,0)}=\left({\rhohat(\mhat,\tau)
s(\mhat,\tau)\over\rhohat(\mhat,0)s(\mhat,0)}\right)^{1/3}
\left[{1+2s(\mhat,\tau)\over 1+2s(\mhat,0)}\right]~.
\label{urho}
\ee
We could use this to find an equation for $s(\mhat,\tau)$ 
explicitly; then
\ba
{\partial\over\partial\tau}\left[{s(\mhat,\tau)e^{8s(\mhat,
\tau)}\over\rhohat(\mhat,\tau)}\right] & = &\nonumber  \left\{{3\qvis(\mhat,\tau)
[1+2s(\mhat,\tau)]\over U(\mhat,\tau)}\right\} \times \\
&&{s(\mhat,\tau)e^{8s(\mhat,\tau)}\over\rhohat(\mhat,\tau)}~,
\ea
with $\uhat(\mhat,\tau)$ evaluated using Eq. (\ref{urho}).
Finally, 
the equation of mass conservation can be written as
\ba
{1\over\rho(M,t)}&=&4\pi r^2(M,t){\partial r(M,t)\over\partial M}
~~\Rightarrow \nonumber \\ {1\over\rhohat(\mhat,\tau)}
&=&4\pi\rhat^2(\mhat,\tau){\partial\rhat(\mhat,\tau)\over\partial
\mhat}~.
\label{massrho}
\ea
If we wish, we can define a new variable $\Vhat(\mhat,\tau)
=1/\rhohat(\mhat,\tau)$, and rewrite the last few equations as
\baray
\Vhat(\mhat,\tau)&=&4\pi\rhat^2(\mhat,\tau){\partial\rhat(\mhat,\tau)
\over\partial\mhat}\nonumber\\
{\uhat(\mhat,\tau)\over\uhat(\mhat,0)}&=&
\left[{\Vhat(\mhat,0)s(\mhat,\tau)\over\Vhat(\mhat,\tau)s(\mhat,0)}
\right]^{1/3}\left[{1+2s(\mhat,\tau)\over 1+2s(\mhat,0)}\right]
\nonumber\\
{\partial\uhat(\mhat,\tau)\over\partial\tau}&=&
- \left ( \frac{a^2}{V(\mhat,\tau)} \left ( \frac{\partial \vhat}{\partial \rhat} \right)^2 \right. \nonumber \\
&& \left. + {2\uhat(\mhat,\tau)[1+s(\mhat,\tau)]\over
3\Vhat(\mhat,\tau)[1+2s(\mhat,\tau)]} \right ) {\partial\Vhat(\mhat,\tau)
\over\partial\tau}
\label{thermov}
\earay
Eqs. (\ref{dynamics}), and either
Eqs. (\ref{firstrho}), (\ref{urho}) and (\ref{massrho}) or Eqs. (\ref
{thermov}), with the initial conditions set up in the previous
sections, can now be cast into finite difference form, with a
suitable specification of the artificial viscous force and heating.

For all calculations done after the code was tested, we also included a Newtonian Gravitation force per unit mass via
\begin{equation}
F_{grav} = - \frac {G M(R) } {R^2},
\end{equation} 
or in difference form,
\begin{equation}
F_j^n = - \frac {G \rho_o \frac{4\pi}{3} (r_j^n)^3} {(R_j^n)^2}.
\end{equation}
This force was added to the conservation of momentum equation, the equation used to set gas velocities. 
Finally, the code self-checks by calculating total energy and momentum to ensure that these are conserved. For energy, the sum of the local energy in each zone is calculated first {\emph{via}}
 
\begin{equation}
E_{\rm kinetic} + E_{\rm therm} = \sum_{i \in zones} ( \frac 1 2 u_{i}^2 + U_{i} ) \Delta M_i.
\end{equation}
The gravitational potential energy is then calculated via
\begin{equation}
E_{\rm grav} = - \sum_{i \in zones} \frac {G M_{enclosed}} {R_i} \Delta M_i .
\end{equation}
and the two energies are added and recorded as the current total energy in the system. Conservation of momentum is also checked though a simple summation:
\begin{equation}
P_{tot} = \sum_{i \in zones} u_{zone} \Delta M_i.
\end{equation}
Finally, the algebraic equation used to determine the total local energy per unit mass of each zone -- the quantity used to determine if a zone was bound or unbound -- was
\begin{eqnarray}
\frac {E_{\rm zone=j}^{tot}} {\Delta M_j} & = & \frac { E_{j}^{therm} + E_j^{kinetic}+ E_{j}^{potential}} {\Delta M_j} \nonumber \\
& = &\frac 1 2 u_{j}^2 + U_j - \frac 1 2 \sum_{i=1}^{i\le j} F_{i}^{grav}  \Delta R_i,
\end{eqnarray}
where $\Delta R_i = R_i - R_{i-1}$.

\section{Sedov Solution}

For the analytic solution to the Sedov problem, we rederived the
solution given in Landau and Lifshitz's \emph{Fluid Mechanics},
thereby finding the correction to an error (in an exponent) that is 
mentioned in Shu's \emph{Gas Dynamics}.
Rather than re-reporting the entire result, we simply note that the correction comes in 
the exponents of the equation for $\rho$:
\be
\rho \propto \left ( \left( \frac {\gamma +1} {\gamma - 1} \right ) \left(1 - {\frac {5t} {2r} v}\right) \right )^{\nu_5}
\ee
where
\[
\nu_5 = - \frac 2 {2 - \gamma}.
\]


\begin{thebibliography}{99}
\bibitem[\protect\citeauthoryear{Chevalier}{1989}]{chev89} Chevalier, R.~A. 1989, ApJ, 346, 847.
\bibitem[\protect\citeauthoryear{Christensen-Dalsgaard}{2003}]{jorgen}Christensen-Dalsgaard, J. 2003, Lecture Notes on Stellar Oscillations, http://astro.phys.au.dk/$\sim$jcd/oscilnotes/, Ch. 5.
\bibitem[Fryer \& Kalogera(2001)]{fry01} Fryer, C.~ L. \& Kalogera, V. 2001, ApJ, 554,548.
\bibitem[\protect\citeauthoryear{Israelian, Rebolo, \& Basri}{Israelian et al.}{1999}]{isr99} Israelian, G., Rebolo, R., \& Basri, G., 1999, Nat, 401, 142.
\bibitem[\protect\citeauthoryear{Landau \& Lifshitz}{1987}]{ll87} Landau, L.~ D., and Lifshitz, E.~ M. 1987, Fluid Mechanics, Pergamon Press: Oxford.
\bibitem[\protect\citeauthoryear{MacFadyen, Woosley \& Heger}{MacFadyen et al.}{1999}]{mac99} MacFadyen, A.~ I., Woosley, S.~ E., \& Heger, A., 1999, ApJ, 524, 262.
\bibitem[\protect\citeauthoryear{MacFadyen, Woosley \& Heger}{MacFadyen et al.}{2001}]{mac01} MacFadyen, A.~I., Woosley, S.~E., \& Heger, A., 2001, ApJ, 550, 410.
\bibitem[\protect\citeauthoryear{Matzner \& McKee}{1999}]{matz99} Matzner, Christopher D. \& McKee, Christopher F. 1999, ApJ, 510,379.
\bibitem[\protect\citeauthoryear{Mirabel et al.}{2002}]{mir02} Mirabel, I.~ F., Mignani, R., Rodrigues, I., Combi, J.~ A., Rodri\'guez, \& L.~F., Guglielmetti, F. 2002, A\&A, 395, 595.
\bibitem[\protect\citeauthoryear{Nadezhin \& Frank-Kamenetskii}{1963}]{nad63} Nadezhin, D.~K. \& Frank-Kamenetskii, D.~A. 1963, Soviet Astronomy, 6, 779.
\bibitem[\protect\citeauthoryear{Orosz \& Bailyn}{1997}]{oro97} Orosz, J. \& Bailyn, C. 1997, ApJ, 477, 876
\bibitem[\protect\citeauthoryear{Richtmyer \& Morton}{1967}]{ric67} Richtmyer, R. \& Morton, K.~W. 1967, Difference Methods For Initial-Value Problems, John Wiley \& Sons: New York
\bibitem[\protect\citeauthoryear{Shahbaz et al.}{1999}]{sha99} Shahbaz, T, van der Hooft, F, Casares, J, Charles, P.~ A., \& van Paradijs, J. 1999, MNRAS, 306, 89
\bibitem[\protect\citeauthoryear{Shu}{1992}]{shu92} Shu, Frank.  1992, The Physics of Astrophysics: Gas Dynamics, Volume II, University Science Books: New York.
\bibitem[\protect\citeauthoryear{Woosley}{1988}]{woos88} Woosley, S.~E. 1988, ApJ, 330, 281.
\bibitem[\protect\citeauthoryear{Woosley \& Weaver}{1995}]{ww95} Woosley, S.~E. \& Weaver. T.~A. 1995, ApJS, 101, 181.
\end{thebibliography}
\end{document}